# Flow dynamics of the Moon

Yu. N. Bratkov

Central Research Institute of Machine Building
Moscow (Korolev), Russia
icm2006 at rambler dot ru

Flow analysis of big basins is given. Internal structure of flows is considered. Correlations between flows are calculated. For example, Mare Orientale is a moving basin. Orientale and Imbrium continental basins are introduced and are considered. Olbers ray crater is a result of precise interaction of the two basins. Flows of the Tycho type are studied. Two Antarctidae, an Indian Ocean, and an America are demonstrated.
Key words: comparative planetology; flow dynamics of solid planet surfaces; streamlined body; shock wave; vortex; moving basin; modelling of superslow planetary flows by high speed flow dynamics.

## I. INTRODUCTION

Comparative planetology was created by the author in [1]. Parallel reading [1] is strongly recommended.

**What we study.** Topology and homological algebra became the main mathematical disciplines during the XX century. (Co)homology is an important global invariant of topological spaces, algebraic systems, etc. It wasn't enough, however, to consider a space as an argument for a cohomology theory. The natural argument for such theories becomes the pair (*space, sheaf*). The structure of a *sheaf* allows to build global objects from local components.

In the middle of the XX century many results of cohomology theory were obtained without any conditions on a space. Under some natural conditions, a space could be reconstructed from the *category of all sheaves* over this space. Such *category of sheaves* with natural axioms (so-called *Grothendieck topos*) is a right object for studying. Thus a primary object is a topos. A space is a secondary object. Topos theory is a 50-years-old algebraic mainstream.

It seems to be obvious to seek and to find some relevant effects in general planetology, global and local. A planet is an image of some generating mapping. The preimage is some nonspacelike object (topos). The existence of some preimage of the Earth was proved in [2] by discrete geometric modelling. The divergence between some *exact structures* on the exact sphere and their physical realizations was found.

So we study the preimage of the Moon. Let it be an exact unit sphere. We consider angular distances on this sphere. Define angular radius of a basin. If we have a circular structure on a planet, we get its radius (in kilometers) and divide it by the radius of the planet (in kilometers). We obtain angular radius of the ring in radians. Distances on a planet are distances along an exact sphere.

Some circles are drawn here on maps of the Moon. We set the center and the radius of the circle under consideration (radius is a distance along the surface of the sphere), and we calculate some points of the circle. These points are drawn on the map. Sometimes points aren't connected by lines.

## II. MARE ORIENTALE

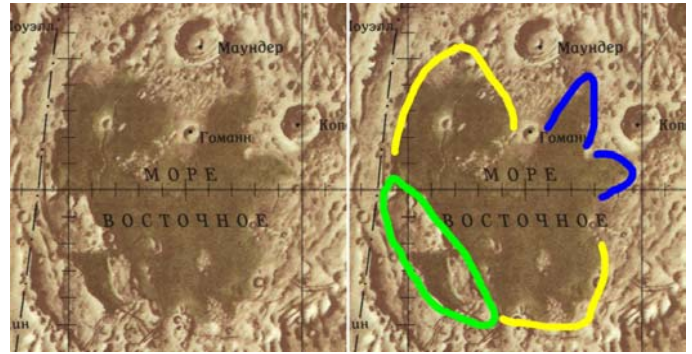

Figure 1. Mare Orientale is a classical flow around a streamlined body. The map is [3]. See [1, p. 18]. Flat front [1, p. 18, Fig. 3, 4], foregoing vortex area, clasters of shock waves, and two back vortices are drawn.

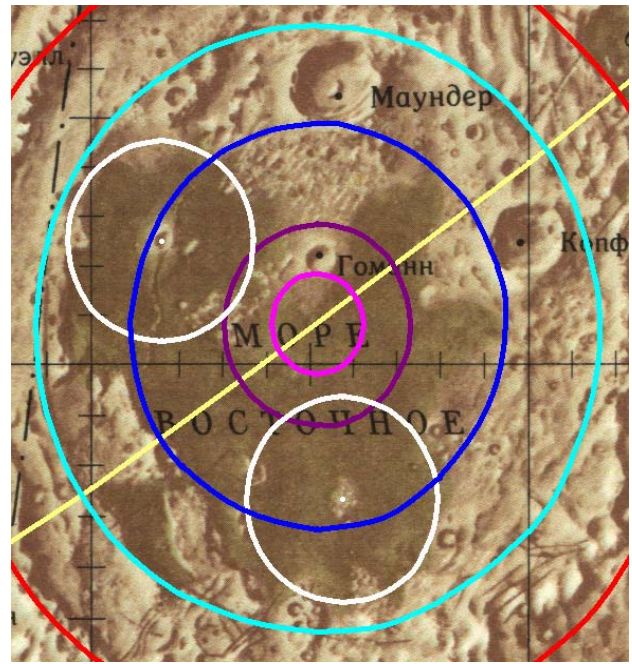

Figure 2. Flow analysis of Mare Orientale. Radii of color circles are (π–3)/8, (π–3)/4, (π–3)/2, (3/4)(π–3) radians. Radii of white circles are (π–3)/4 radians, centers are pointed. Yellow straight line connects centers of Mare Orientale and Mare Imbrium.

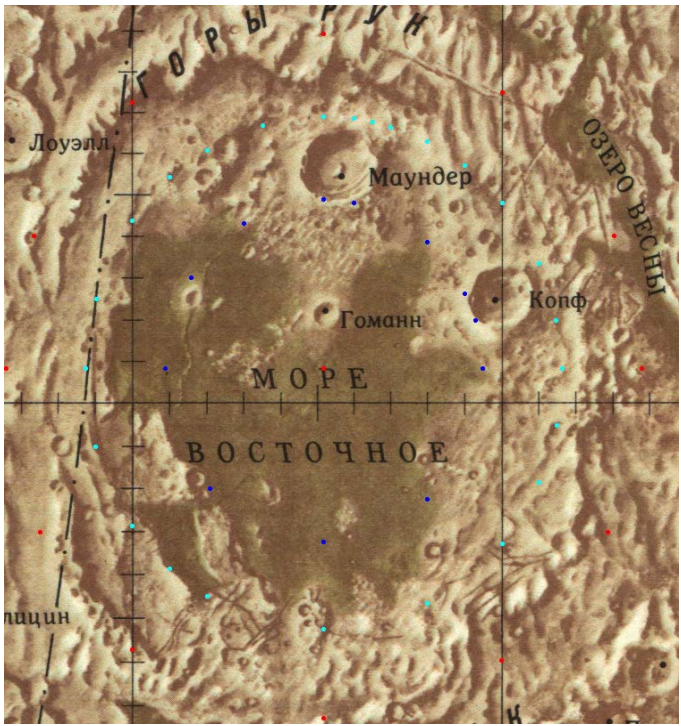

Figure 3. Pointwise circles allow to see Lunar relief. The foregoing vortex area (see Fig. 1) is placed exactly between the blue and cyan circles. One could see that the cyan circle goes along some rock circle with good precision. This rock circle becomes apparent after this eigenvalue analysis.

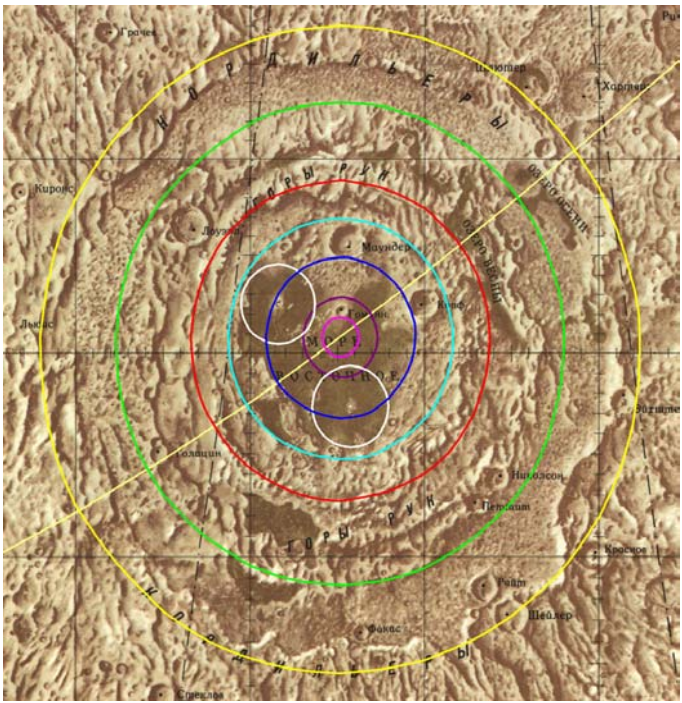

Figure 4. Flow analysis of Mare Orientale. Radii of additional circles are (π–3), (3/2)(π–3), 2(π–3) radians. The center is (19.17⁰ S, 94.82⁰ W).

The blue circle is tangent to Maunder and Kopff craters, so these craters are two back vortices of blue moving basin. We see here an example of multilevel flow structure [1, pp. 16–17]. The first level, with blue back vortices, is shown in Figure 1.

The smallest circle is tangent to Hohmann crater. The picture is similar to Maunder and Kopff craters, but here is one crater only. It is interesting to understand what element of the flow it is. See, for example, [1, p. 14, Fig. 5].

White circles represent shock waves. Such big round vortices as shock waves is an interesting Lunar phenomenon (see [1, p. 29, Fig. 4]). Centers of white circles are symmetric with respect to the blue circle. The upper white circle is centered in Il'in crater. Maybe this circle must be bigger. However, one could see that Il'in crater isn't impact too.

The conclusion is: Maunder, Kopff, Hohmann, Il'in craters aren't impact. They are elements of the flow. Oriental Basin obviously isn't impact too.

See general theory of eigenvalues of planetary flows in [1, pp. 8, 37, 45].

Recall that the existence of discrete spectrum of a bounded body is wellknown (at least for creators of quantum mechanics). The Moon is a bounded body.

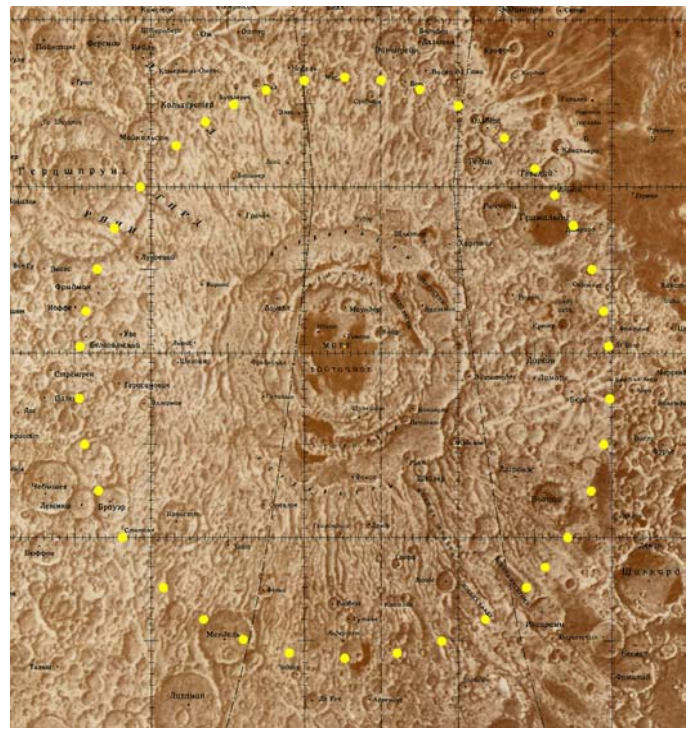

Figure 5. Continental basin of Mare Orientale (see [1, p. 45, Fig. 1–4]). The map is [4]. Radius of the circle is 4(π–3) radians. The basin's continental size is Earthlike [1, p. 45, Fig. 1–4].

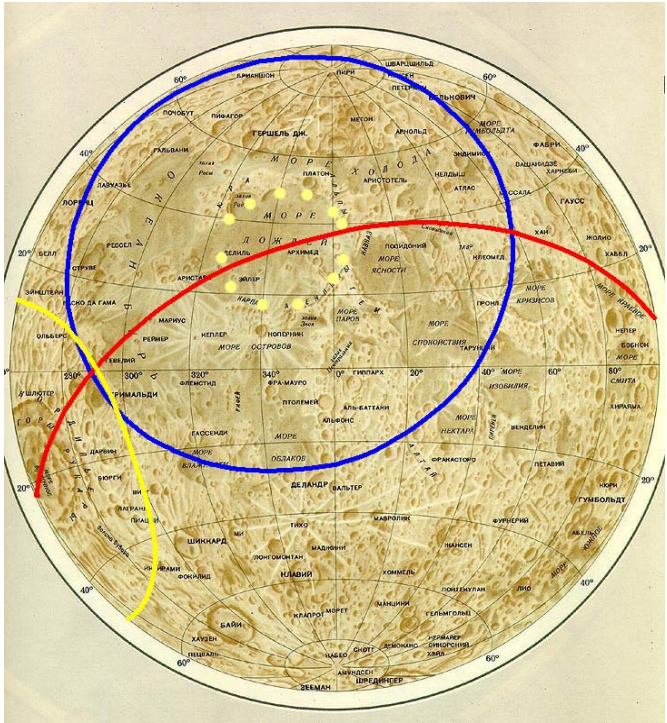

Figure 6. Compare Oceanus Procellarum and Fig. 1. Continental basins of Mare Orientale and Mare Imbrium are drawn. Straight line connects their centers. The map is [5]. Radii of circles are $2(\pi-3)$ (the center is ($33.25^0$ N, $15.82^0$ W)), $4(\pi-3)$, and $(\pi-1)/2$ radians (Earthlike and Marslike continental sizes [1, p. 45, Fig. 1–4; p. 46, Fig. 6]).

### III. OLBERS—GLUSHKO RAY SYSTEM ON THE BORDER OF ORIENTAL AND IMBRIUM CONTINENTAL BASINS

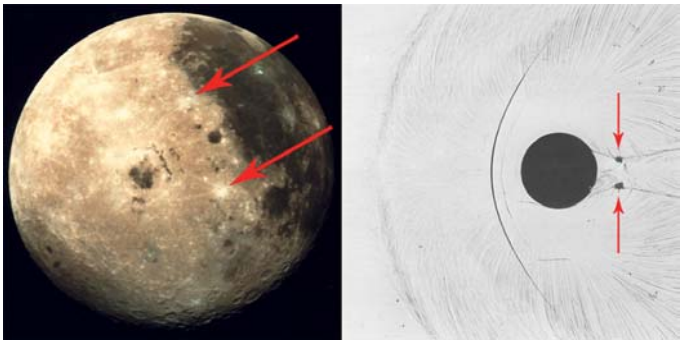

Figure 7. Back vortices of Oriental Basin. Olbers—Glushko is the upper vortex. **Left:** Image: Galileo, P-37329. **Right:** The cylinder is streamlined by supersonic flow [6, Introduction]. M = 2.5, Re = 735 000.

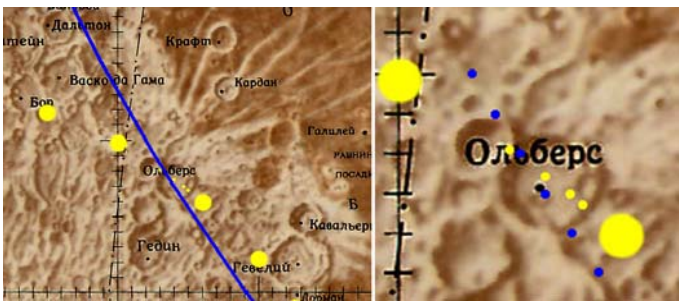

Figure 8. Fragments of Fig. 5. The border of the continental basin of Mare Imbrium (see Fig. 6) is added. **Right:** Points of the two circles were calculated and drawn with good precision. The intersection point is between Olbers (right) and Glushko (left) craters.

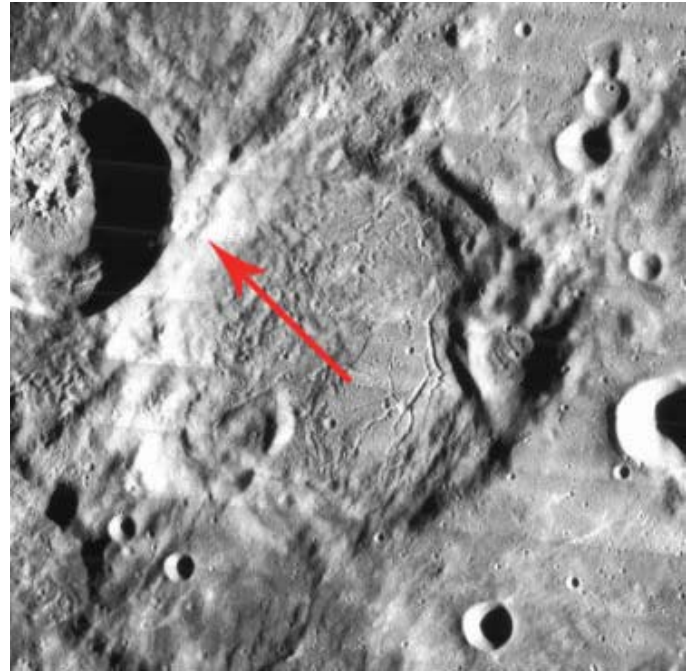

Figure 9. Image: Lunar Orbiter, LO-IV-174H_LTVT. There is a crater at the intersection point (Fig. 8). It is the main source of rays (Fig. 8).

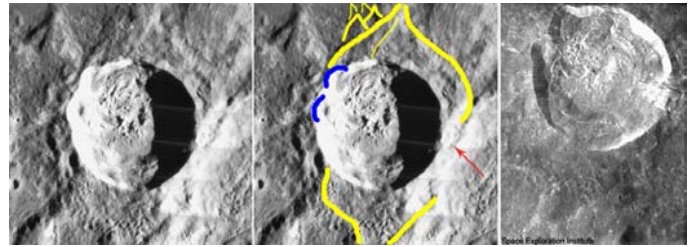

Figure 10. Glushko crater is a moving basin. One could see two back vortices and side shock waves. See [1, p. 16, Fig. 3; p. 18]. Olbers has side shock waves too (Fig. 9). Glushko and Olbers are moving in the same direction. It means that the intersection point (Fig. 8) is moving too, in the same direction. Thus the distance between Orientale and Imbrium is rising (see Fig. 6). **Left:** LO-IV-174H_LTVT. **Right:** SMART-1 (August 2005).

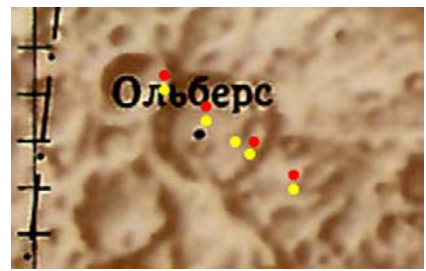

Figure 11. Fragment of Fig. 5. Red points are added. Yellow points are points of the circle with radius $4(\pi-3)$ radians, and red points are points of the circle with radius $(\pi-2)/2$ radians. In [1, p. 45] the question of true continental diameter ($8(\pi-3)$ or $\pi-2$) is a difficult unanswered question. There is small difference for radii $4(\pi-3)$ and $(\pi-2)/2$, it is 28 km on the Earth surface. We have an answer on the Moon: the true continental diameter is $8(\pi-3)$ radians. The yellow circle is the center line of Olbers crater with high precision. Therefore Olbers crater is the main body of this source of rays.

## IV. THE BORDER OF MARE IMBRIUM'S CONTINENTAL BASIN

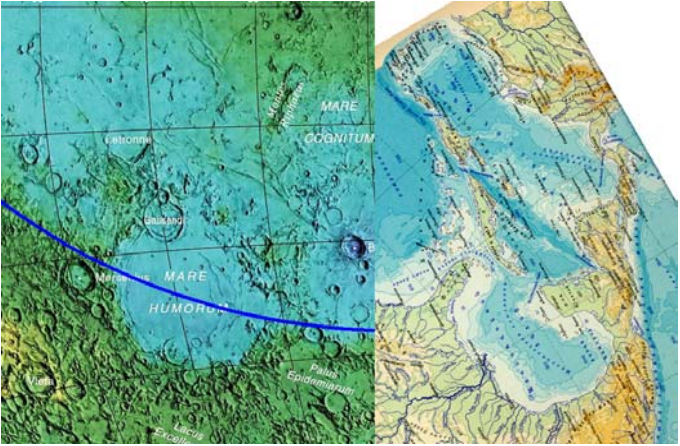

Figure 12. Mare Humorum (left, the map is [7]) is a copy of Caribbean Antarctida (right, the map is [8, p. 191]). See [1, p. 24, Fig. 7, 10]. The blue line is the border of the continental basin of Mare Imbrium (Fig. 6). This line goes through the center of Mare Humorum. See [1, p. 46, Fig. 6].

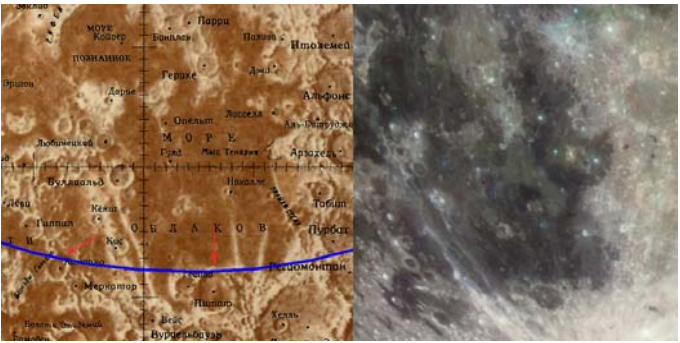

Figure 13. The blue line is tangent to Mare Nubium and to Campanus (left arrow) and Pitatus (right arrow) craters. Mare Nubium's body and Pitatus crater are separated by this line. **Left:** [4]. **Right:** PIA00405.

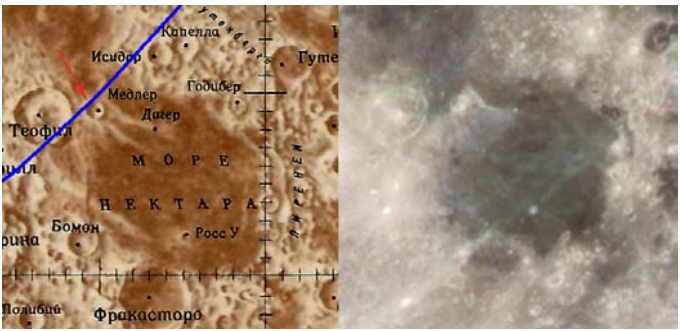

Figure 14. The blue line is tangent to Mare Nectaris and to Mädler crater (arrow). **Left:** [4]. **Right:** PIA00405.

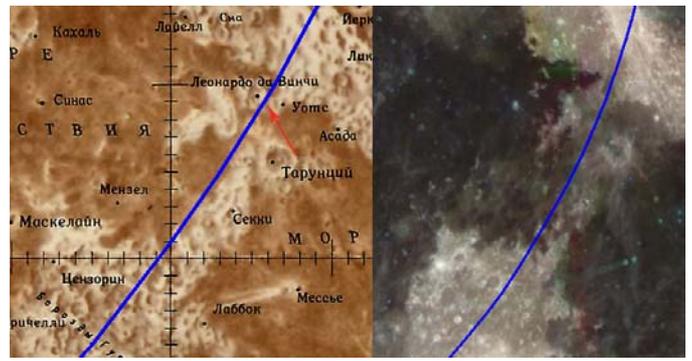

Figure 15. The blue line is tangent to Mare Tranquillitatis ant to da Vinchi crater (arrow). Mare Tranquillitatis and Mare Fecunditatis are separated by this line. **Left:** [4]. **Right:** PIA00405.

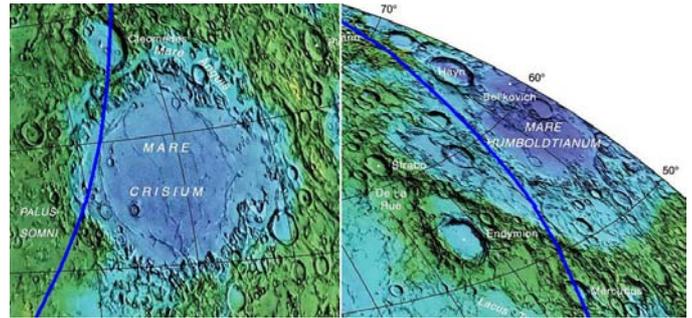

Figure 16. The blue line is tangent to internal basins of Mare Crisium and Mare Humboldtianum. The map is [7].

Note that the blue line is tangent to some middle size craters (see Fig. 13–15). There aren't too many tangent craters (of middle size) outside these tangent seas (the author found Glushko (Fig. 8) and Nernst, and no more), but each tangent sea (Fig. 13–15) has its own embedded tangent crater or two.

## V. MARE CRISIUM

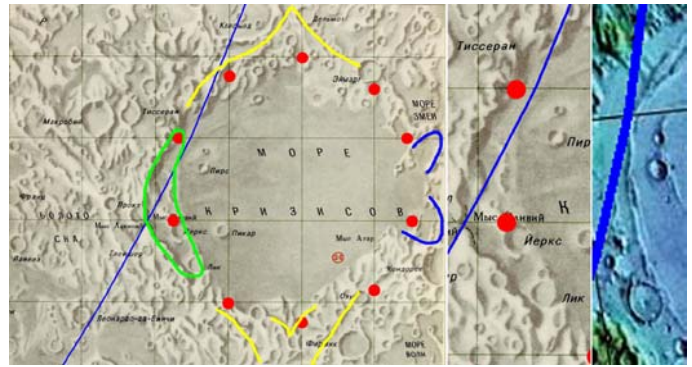

Figure 17. Mare Crisium is a moving basin [1, p. 18, 16]. Radius of the circle is $\pi-3$ radians; its center was taken roughly. One could see foregoing area, flat front (see the central image) [1, p. 18, Fig. 3–5, 12], two back vortices, side shock waves. The map is [8, p. 7]. See Fig. 16 left.

See Fig. 18. The center $c_2$ of left family of concentric circles *is defined by two conditions*. 1) The distance to the center of Proclus ray crater is $\pi-3$ radians. 2) The distance to the center of Imbrium basin is $(\pi-1)/2 + (\pi-3)/2 = \pi-2$ radians. By the 2-nd condition, the light blue circle is tangent to the blue line.

The yellow straight line connects the center $c_1$ of Proclus ray crater and the center $c_2$ of the light blue circle. *By definition*, the center $c_3$ of the red brown circle is an intersection point of the yellow straight line and the light blue circle. Therefore, the distance between $c_1$ and $c_3$ is $(3/2)(\pi-3)$ radians.

Let the center $c_1$ of Proclus ray crater be (15.952 N, 46.982 E). Then $c_2$ = (16.786 N, 55.385 E), $c_3$ = (17.083 N, 59.615 E). The distance from $c_2$ to the center of Byrgius A ray crater (Fig. 7, lower arrow) is $\pi-1$ radians exactly. Thus Proclus and Byrgius A ray craters aren't impact.

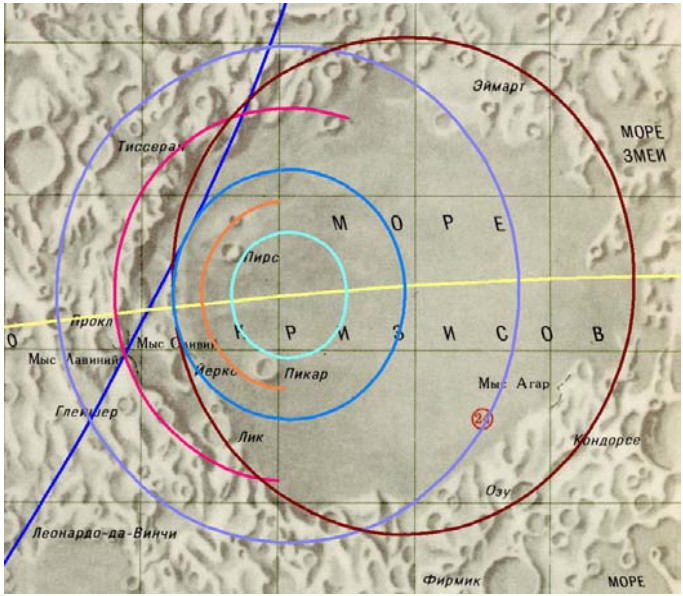

Figure 18. Analysis of Mare Crisium. Radii of circles are $(\pi-3)/4$, $(3/8)(\pi-3)$, $(\pi-3)/2$, $(3/4)(\pi-3)$, $\pi-3$, $\pi-3$ radians. Many craters are tangent or centered with respect to the given circles. It isn't easy to find here an impact crater. Peirce and Picard craters (between lightest blue and orange circles) are analogous to the centers of white circles of Mare Orientale (Fig. 2).

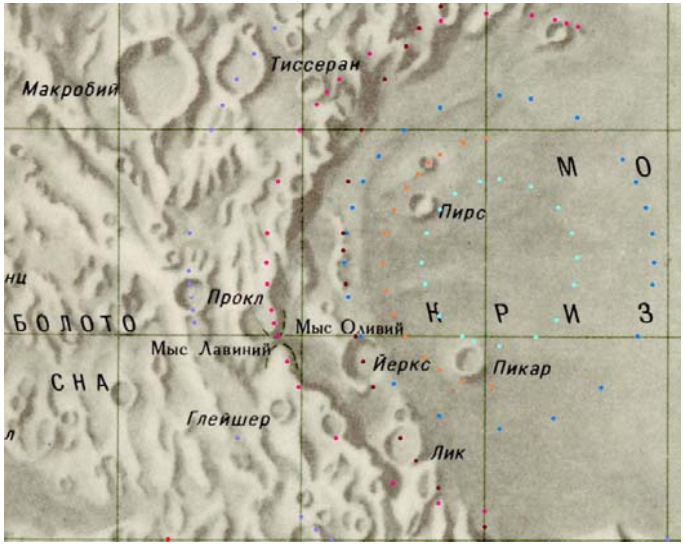

Figure 19. Fragment of Fig. 18. Pointwise circles allow to see details of Lunar relief. The foregoing area (Fig. 17) is placed exactly between light blue and pink circles. Compare rays of Proclus (Fig. 19), Stevinus (Fig. 21), Eta Carinae Nebula (Fig. 20 lower).

## VI. RAY FLOWS OF THE TYCHO TYPE

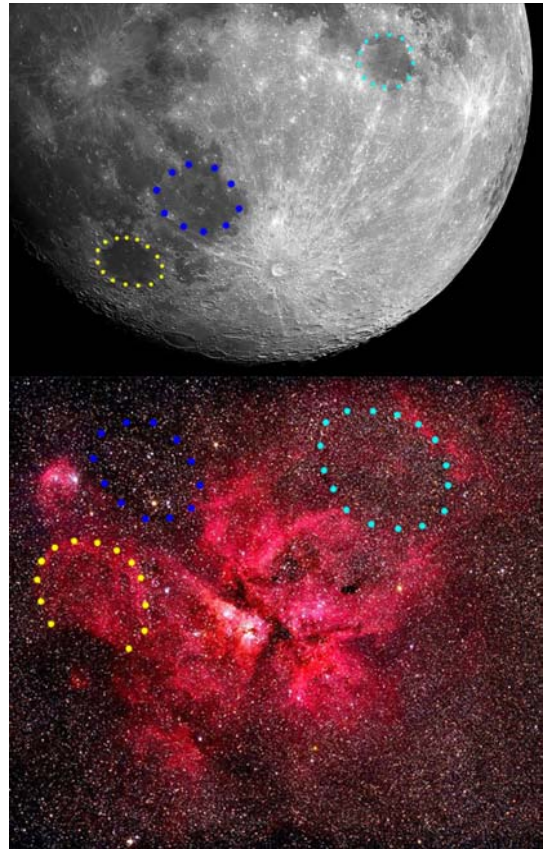

Figure 20. Tycho ray crater, Mare Humorum, Mare Nubium, Mare Nectaris in Eta Carinae Nebula. Images: [9]. Comparing: Yu. N. Bratkov.

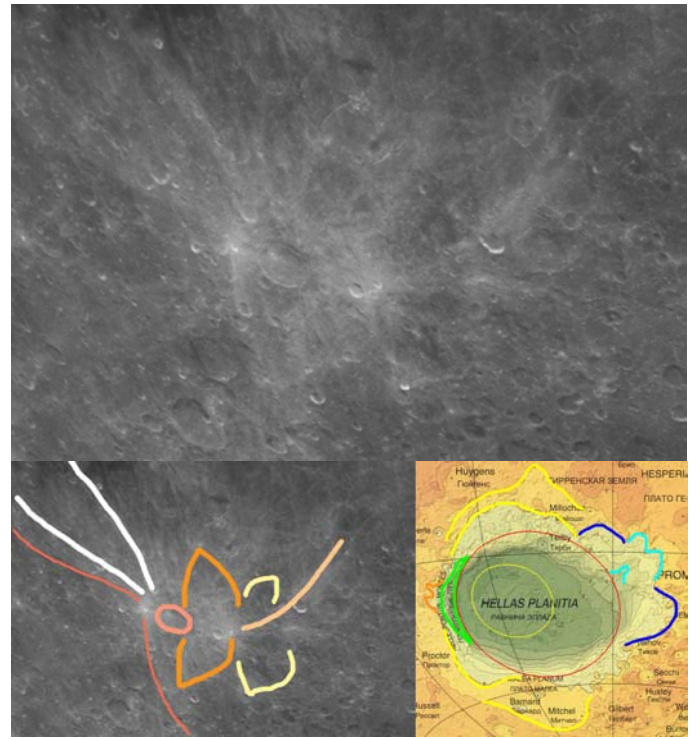

Figure 21. **Upper:** Ray crater Stevinus is a tail of the Tycho flow (Fig. 20, upper right corner). Stevinus flow is a copy of Tycho flow, and it has its own

small Stevinus in the tail. Compare with Fig. 19: orientation of Proclus' rays is the same. Image: PIA00303. **Lower left:** Colored Stevinus. **Lower right:** Mars. Hellas Planitia is a moving basin [1, p. 22, Fig. 6]. Foregoing whiskers, wings, and oscillating tail are analogous to Stevinus' ones.

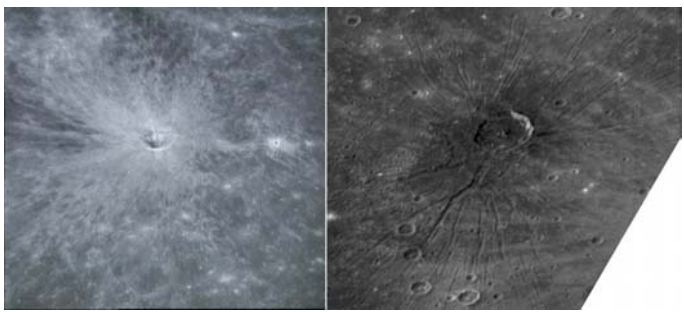

Figure 22. Examples of ray flows, see Fig. 20, 21. **Left:** The Moon. An unnamed ray crater. The center of the picture is ($4^0 30'$ N, $100^0$ E). A Stevinus and a Mare Nectaris in the tail are being. Image NASA: AS11-42-6285. **Right:** Mercury. A convex down ray flow. See [1, p. 6, Fig. 1]. PIA10934 (mirror image).

## VII. MARE IMBRIUM AND AROUND

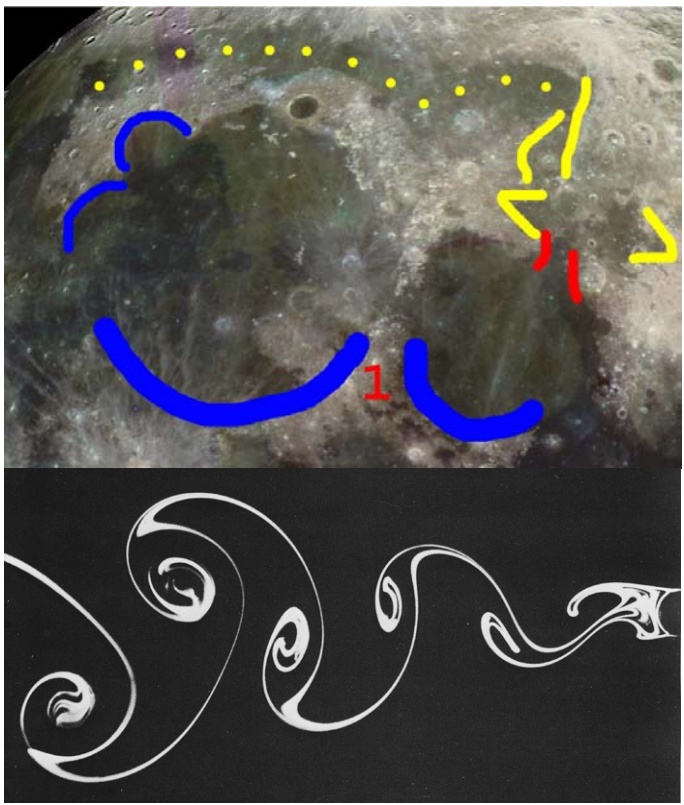

Figure 23. **Upper:** One more Lunar Antarctida (see Fig. 12) (it is a convex down flow) [1, p. 28, Fig. 9]. It was introduced in [1] as a 3-component Antarctida (usually they are 2-component). Lacus Somniorum is a head of the flow. Mare Frigoris is a long shock wave of the head. Such shock wave usually connects a head with a big continent (Procellarum in this case; we don't distinguish convex up and convex down continents here). PIA00405. **Lower:** Oscillating trace in water behind streamlined cylinder. Re = 140. [6, ph. 94].

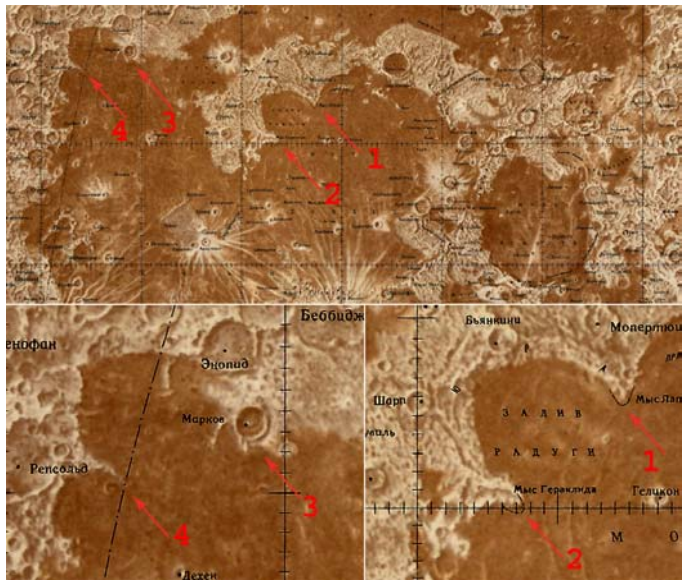

Figure 24. Now we can see the 4-th component of the Somniorum-Serenitatis-Imbrium flow. It is the Northern part of Oceanus Procellarum. 1 = 3, 2 = 4. The map is [4].

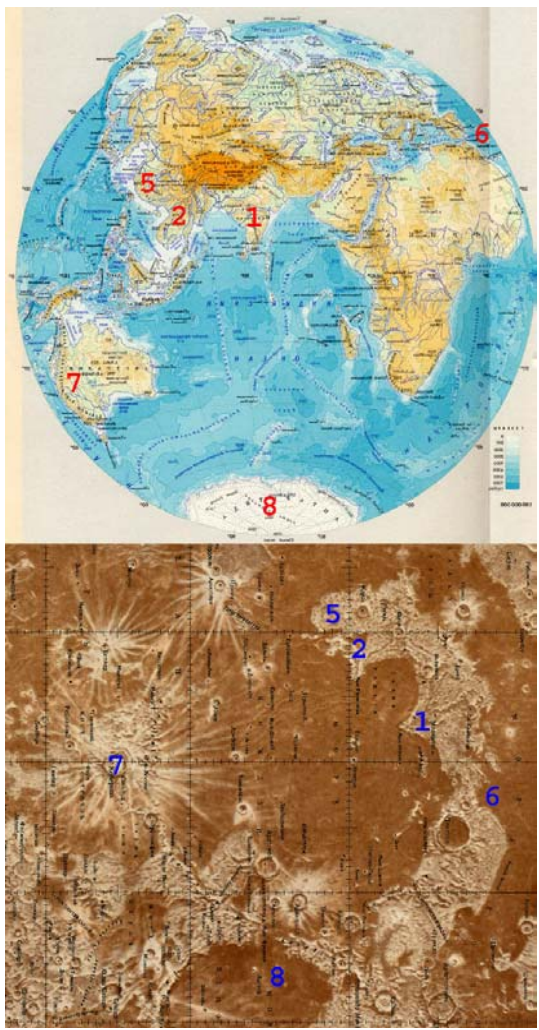

Figure 25. By [1, p. 24, Fig. 10], Imbrium is an Indian Ocean. However, Imbrium Indochina isn't mirrored. Maps are [8. p. 17], [4] (north is right).

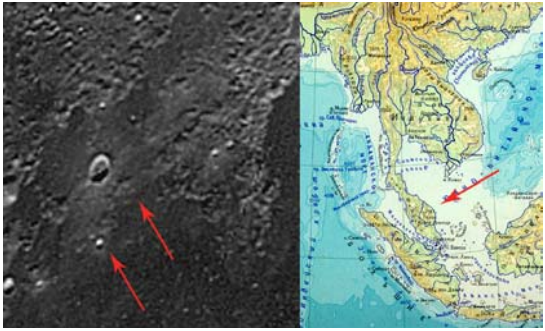

Figure 26. **Left:** Imbrium Malacca. PIA00128. **Right:** Earthen Indochinese Malacca [8, p. 131]. Malaccae are necessary for Indochinae, see [1, p. 12; p. 33, Fig. 2, 6].

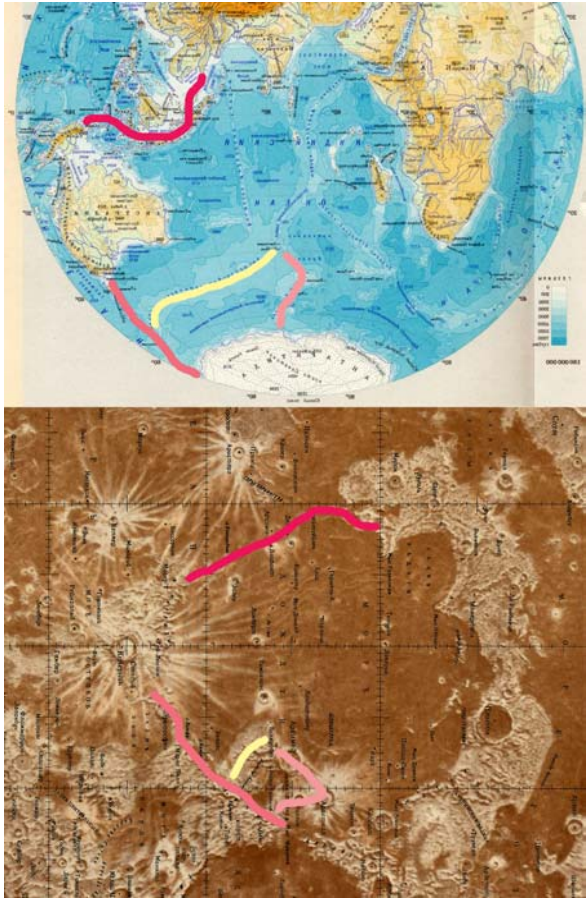

Figure 27. Comparing ranges. See Fig. 25, 26. **Lower:** North is right.

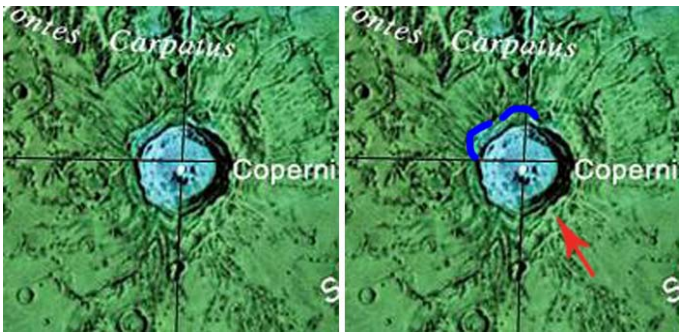

Figure 28. Copernicus is a moving basin. One could see two back vortices (see Fig. 7 right, [1, p. 18, Fig. 1, 2, 4 left]), flat front (arrow, see [1, p. 18,

Fig. 3–5, 12]). The direction of moving is contrary to the arrow. By Fig. 25, 27, Copernicus is analogous to the moving basin of Australia (see [1, p. 16, Fig. 3, 9]). Copernicus ray crater isn't impact. The map is [7], north is up.

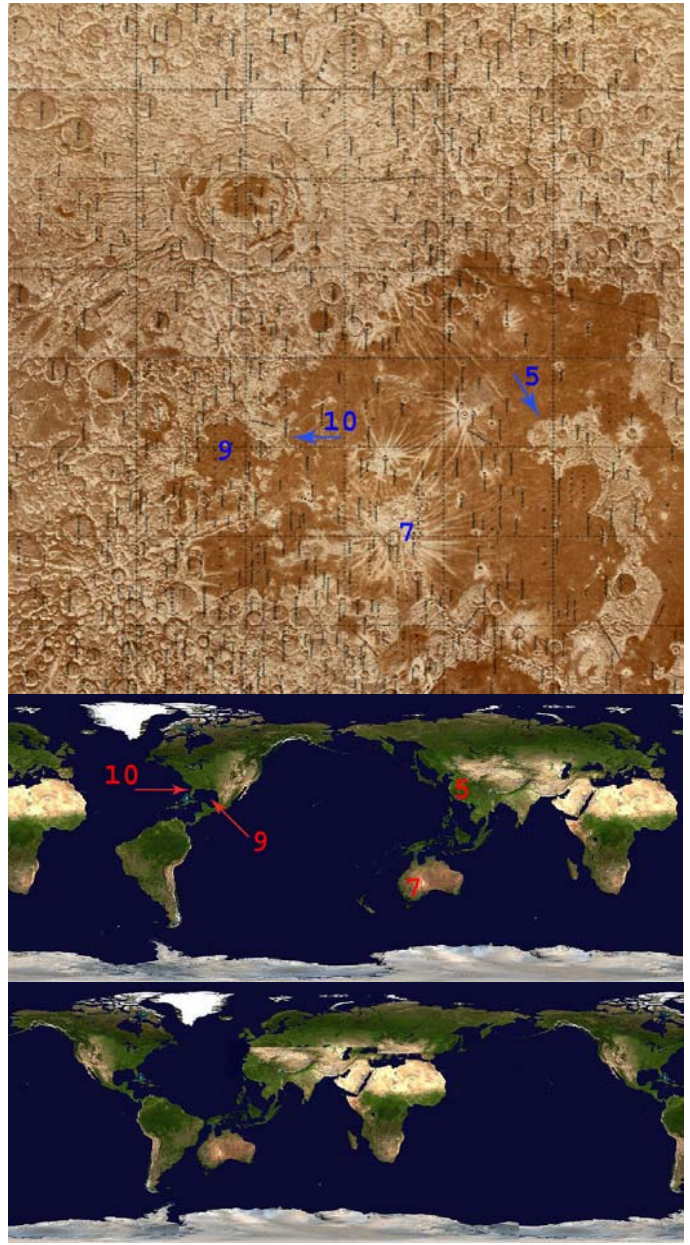

Figure 29. By Fig. 12, Mare Humorum is a Gulf of Mexico. However, Lunar North America isn't mirrored, so Procellarum is an Atlantic Pacific. Western coast of Procellarum (upper) is an Atlantic Coast. See [1, p. 5, Fig. 3]. Let South Pole—Aitken basin be a lower right big basin. **Upper:** North is right.

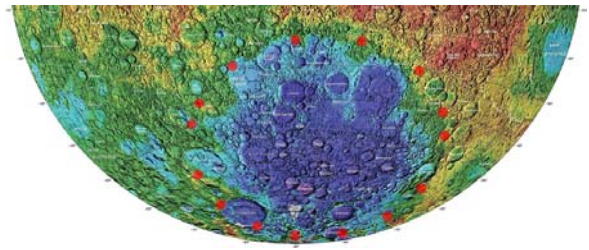

Figure 30. South Pole—Aitken basin has Earthlike continental size [1, p. 45, Fig. 1-4]. Radius of the circle is $4(\pi-3)$ radians. The map is [7].

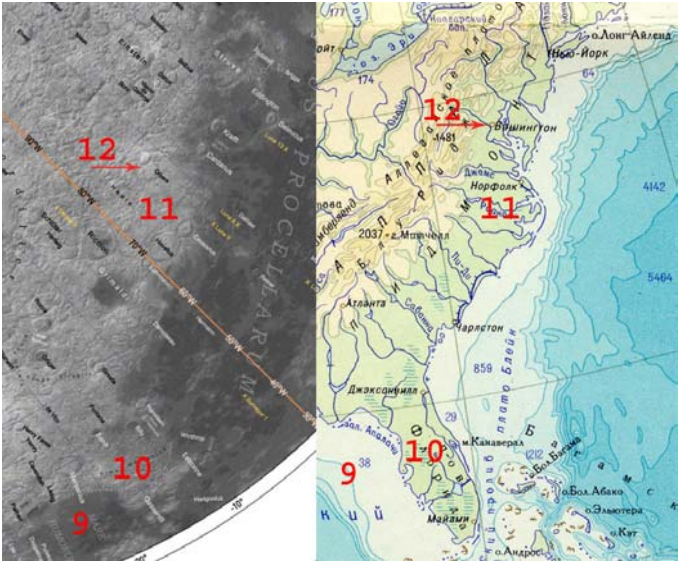

Figure 31. Comparing Atlantic coast-lines (Fig. 29, 12). **Left:** 12 is Olbers—Glushko ray system, see Fig. 7─10. Glushko is upper. The map is [10]. **Right:** 12 is Washington. The map is [8, p. 191].

## VIII. Mascons

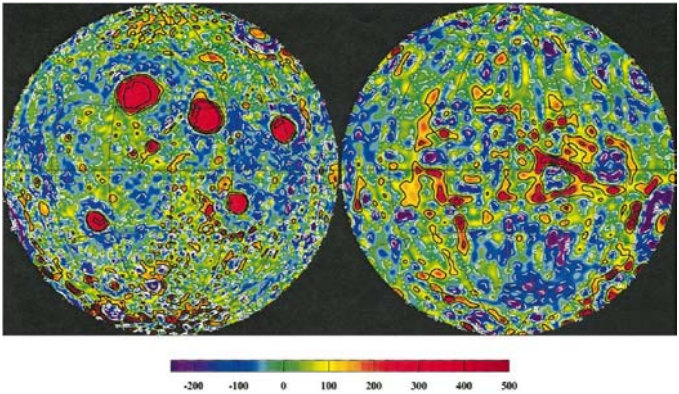

Figure 32. Lunar gravity map [11, Fig. 8].

By [1], mascons (Fig. 32) are convex down flows. They are analogous to cold atmospheric flows. Cold flows are more dense, so they move down. Hot flows are of less density, so they move up. Thus, convex down flows (bottoms of seas and oceans) are flows of high density. Obviously they are mascons themselves. See [1, p. 21] for understanding their nature.

Nevertheless, flows of small density (mountains) can have big volume, so they can be mascons too. Analogously, giant South Pole—Aitken basin (a convex down flow) isn't a mascon.